\documentclass{webofc}
\usepackage[varg]{txfonts}   
\begin{document}
\title{Strange mesons in nuclei and neutron stars}
%
%

\author{\firstname{Laura} \lastname{Tolos}\inst{1,2,3}\fnsep\thanks{\email{tolos@ice.csic.es}} }

\institute{ Institute of Space Sciences (ICE, CSIC), Campus UAB, Carrer de Can Magrans, 08193, Barcelona, Spain
\and
           Institut d'Estudis Espacials de Catalunya (IEEC), 08034 Barcelona, Spain
\and
           Frankfurt Institute for Advanced Studies, Ruth-Moufang-Str. 1, 60438 Frankfurt am Main, Germany
          }

\abstract{%
The present status in the field of strange mesons in nuclei and neutron stars is reviewed. In particular, the $\bar K N$ interaction, that is governed by the presence of the $\Lambda(1405)$, is analyzed and the formation of the $\bar K NN$ bound state is discussed. Moreover, the properties of $\bar K$ in dense nuclear matter are studied, in connection with strangeness production in nuclear collisions and kaon condensation in neutron stars.}
\maketitle
\section{Introduction}
\label{intro}

Understanding the dynamics of hadrons with strangeness has received a lot attention over the past decades in connection with the study of exotic atoms, the analysis of strangeness production in particle and nuclear research facilities, and the investigation of different strange phases in the interior of neutron stars (for a recent review see Ref.~\cite{Tolos:2020aln}). 
In particular, one venue of interest in the field of strangeness is the study of strange mesons, such as the $\bar K$ meson, and its dynamics with nucleons and nuclear matter.

It is known that the behaviour of the $\bar KN$ scattering amplitude close to threshold is governed by the presence of the $\Lambda(1405)$ resonance below threshold. The nature of this baryonic resonant state has been carefully studied, in particular after the theoretical predictions that describe this state as the superposition of two poles \cite{Oller:2000fj,Jido:2003cb,Hyodo:2007jq}. Furthermore, the two-pole nature of the $\Lambda(1405)$ shows that the $\bar K N$ interaction could be attractive enough to produce bound states, such as the $\bar KNN$ state (see Ref.~\cite{Nagae:2016cbm} and references therein).

Also, the properties of $\bar K$  in dense nuclear matter have prompted the interest of the scientific community, in particular after the theoretical predictions of kaon condensation in neutron stars \cite{Kaplan:1986yq}, and the analysis of strangeness creation and propagation in nuclear collisions  (see, for example, \cite{Fuchs:2005zg,Hartnack:2011cn}). 

In this paper we review the  status in the field of strange mesons in nuclei and neutron stars, paying a special attention to the formation of the $\bar KNN$ state, the propagation of strange mesons in nuclear collisions and the phenomena of kaon condensation in neutron stars.

\section{The $\bar KN$ interaction: the $\Lambda(1405)$}
\label{lambda1405}
The $\bar K N$ scattering in the $I=0$ channel is governed by the presence of the $J^P=1/2^-$ $S=-1$ $\Lambda(1405)$. The origin of the $\Lambda(1405)$ as a molecule was predicted more than 50 years ago by Dalitz and Tuan  \cite{Dalitz:1959dn,Dalitz:1959dq}. And, since the end of the last century, a lot of effort has been invested in understanding the $\bar K N$ interaction and the nature of the $\Lambda(1405)$ using coupled-channel unitarized theories within meson-exchange models \cite{MuellerGroeling:1990cw,Haidenbauer:2010ch} or meson-baryon chiral effective theories ($\chi EFT$) \cite{Kaiser:1995eg, Oset:1997it,Oller:2000fj,Lutz:2001yb,GarciaRecio:2002td,Jido:2003cb,Borasoy:2005ie,Oller:2006jw,Feijoo:2018den}. A very interesting conclusion of these works is that the dynamics of the $\Lambda(1405)$ is described by the superposition of two poles of the scattering matrix, between the $\bar K N$ and $\pi \Sigma$ thresholds \cite{Oller:2000fj,Jido:2003cb,Hyodo:2007jq}. Whereas experimentally the $\Lambda(1405)$ is seen as one resonance, the existence of two poles is supported in  reaction-dependent line shapes \cite{Jido:2003cb}.

\section{The $\bar KNN$ bound state}
\label{knn}

\begin{table}[]
    \centering
    \begin{tabular}{c|c|c|c|c}
        {\rm Reference} & {\rm B [MeV]} & {$\Gamma$ [MeV]}  & {\rm Method} & {\rm Potential}\\
         \hline
         \hline
         {\rm Barnea et al. (2012)} & 16 & 41 & {\rm Variational} & {\rm chiral} \\
         \hline
          {\rm Dote et al.  (2009)} & 17-23 & 40-70 & {\rm Variational } & {\rm chiral} \\
         \hline
          {\rm Dote et al.  (2018)} & 14-50 & 16-38 & {\rm ccCSM } & {\rm chiral} \\
         \hline
          {\rm Ikeda et al. (2010) } & 9-16 & 34-46 & {\rm Faddeev } & {\rm chiral} \\
         \hline
          {\rm Bayar et al.  (2013) } & 15-30 & 75-80 & {\rm Faddeev } & {\rm chiral} \\
           \hline
          {\rm Sekihara et al. (2016) } & 15-20 & 70-80 & {\rm Faddeev } & {\rm chiral} \\
          \hline
          {\rm Yamazaki et al. (2002)} & 48 & 61 & {\rm Variational } & {\rm phenomenological} \\
         \hline
          {\rm Shevchenko et al. (2007)}  & 50-70 & 90-110 & {\rm Faddeev} & {\rm phenomenological} \\
         \hline 
         {\rm Ikeda et al. (2007)}  & 60-95 & 45-80 & {\rm Faddeev } & {\rm phenomenological} \\
         \hline
          {\rm Wycech et al. (2009) } & 40-80 & 40-85 & {\rm Variational } & {\rm phenomenological} \\
          \hline
          {\rm Dote et al. (2017) } & 51 & 32 & {\rm ccCSM  } & {\rm phenomenological} \\
          \hline
           {\rm Revai et al. (2014) } & 32/ 47-54 & 50-65 & {\rm Faddeev} & {\rm chiral/phenomenological} \\
        \end{tabular}
    \caption{Binding energy and width of $K^-pp$ for different chiral and phenomenological calculations using variational, Faddeev or ccCSM+Feshbach methods. Table modified from Table 2 of Ref.~\cite{Tolos:2020aln}}
    \label{tab:KNN}
\end{table}

The two-pole structure of the $\Lambda(1405)$ indicates that the $\bar K N$ interaction might be attractive enough to produce bound states.  As a consequence, $\bar K$-nuclear clusters may form, such as the $\bar K NN$ in isospin $I=1/2$ and $J^{\pi}=0^-$. Indeed, the $I=1/2$ $\bar K NN$ state has been extensively studied over the last decades, both theoretically and experimentally (see Ref.~\cite{Nagae:2016cbm} and references therein). 

Several theoretical works have addressed the existence of this state. In Table \ref{tab:KNN}  the  $K^- pp$ binding energies $(B)$ and widths $(\Gamma)$ resulting from different models are shown (see references in Ref.~\cite{Tolos:2020aln}). In this table, we separate between variational, three-body Faddeev calculations or the coupled-channel Complex Scaling Method (ccCSM), whereas, at the same time, we distinguish between those models where the two-body interactions are based on $\chi EFT$ or energy-independent phenomenological models. We observe that the binding energies range between 9 and 95 MeV, while the decay widths cover values from 16 to 110 MeV. 

The variety of values for the binding energy and width is due to different sources: a) the uncertainties in the subthreshold extrapolation of the $\bar KN$ interaction. The $\bar KN$ interaction from $\chi EFT$ is less attractive than the phenomenological potentials in the subthreshold region. Hence, the chiral interactions give binding energies considerably lower than those from  phenomenological interactions; b) the use of variational or Faddeev calculations introduces certain approximations. Full three-body interactions are not accounted for in variational methods, whereas Faddeev calculations deal with separable two-body interactions. As for ccCSM, it combines the merits of variational and Faddeev but with a high computational cost.

\section{Antikaons in matter}
\label{matter}

Antikaons in matter have been extensively studied over the last decades. Early works on relativistic mean-field models (RMF)  \cite{Schaffner:1996kv} or quark-meson coupling (QMC) schemes \cite{Tsushima:1997df} obtained potentials of a few hundreds of MeVs at saturation density $\rho_0$. However, most of these models  assumed that the low-density theorem is valid for the calculation of the $\bar K$ optical potential in dense matter, which is doubtful due to the presence of the $\Lambda(1405)$ close to threshold. Thus, it is of fundamental importance to study the properties of the $\Lambda(1405)$ in matter and the effect on the $\bar K N$ scattering amplitude and, hence, on the $\bar K$ optical potential (or self-energy).

One way to proceed is to use unitarized theories in coupled channels in dense matter, using $\chi EFT$ with strangeness \cite{Waas:1996fy,Lutz:1997wt,Ramos:1999ku} or from meson-exchange models \cite{Tolos:2000fj,Tolos:2002ud}. Within these approaches, an attractive antikaon potential  with density is obtained due to the modified $S$-wave $\Lambda(1405)$ resonance in the medium. This modification emerges from different effects: a) the Pauli blocking on the baryon states in the intermediate meson-baryon propagator \cite{Koch:1994mj}; b) the inclusion of the $\bar K$ self-energy \cite{Lutz:1997wt} in the propagation of $\bar K$ in matter; and  c) the implementation of self-energies of the other mesons and baryons in the intermediate states \cite{Ramos:1999ku}. 

\begin{figure*}[t]
  \centering
 \includegraphics[width=0.7\textwidth]{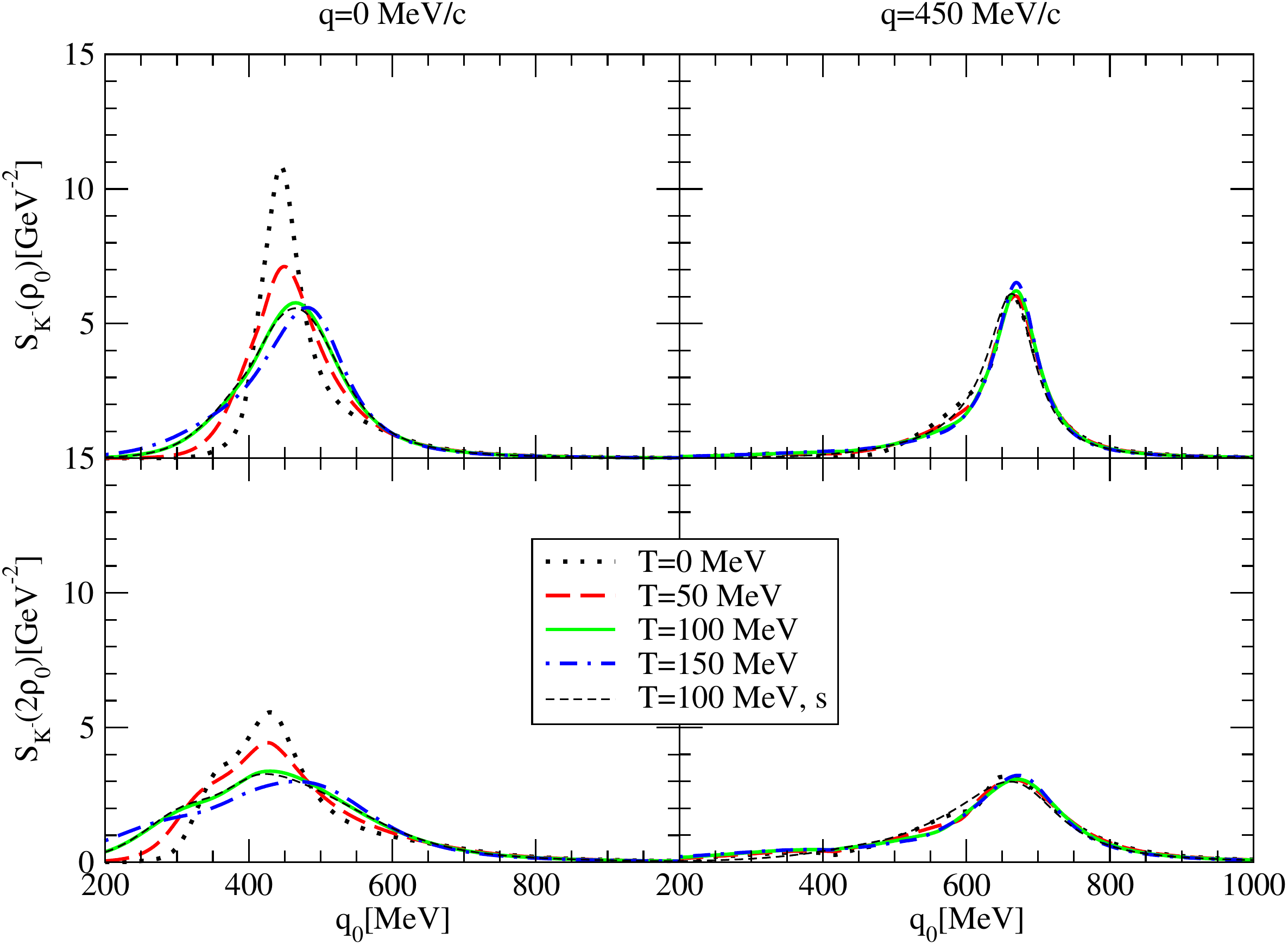}
\caption{$\bar K$ spectral function for different momenta, densities and temperatures from the chiral model of Ref.~\cite{Tolos:2008di}. }
\label{Kbarspectral}
\end{figure*}

Once the $\bar K N$ effective interaction in matter and, hence, the $\bar K$ potential (and self-energy) are determined, it is possible to obtain the spectral representation of the $\bar K$ propagator as
\begin{equation}
S_{\bar K} (\omega,{\vec q})= -\frac{1}{\pi} {\rm Im}\, D_{\bar K}(\omega,{\vec q})
= -\frac{1}{\pi}\frac{{\rm Im}\, \Pi_{\bar K}(\omega,\vec{q})}{\mid
\omega^2-\vec{q}\,^2-m_{\bar K}^2- \Pi_{\bar K}(\omega,\vec{q}) \mid^2 } \ ,
\label{eq:spec}
\end{equation}
where $D_{\bar K}(\omega,{\vec q}\,)$ stands for the $\bar K$ propagator and $\Pi_{\bar K}(\omega,\vec{q})$ is the $\bar K$ self-energy. Note that $\omega$ is the energy, $\vec{q}$ the momentum  and $m_{\bar K}$ the mass of the $\bar K$ meson.

The $\bar K$ spectral function for the chiral model of Ref.~\cite{Tolos:2008di} is shown in Fig.~\ref{Kbarspectral}. The $\bar K$ spectral function considers the $S$- and $P$-wave contributions for different temperatures (different lines) for two densities ($\rho_0$ in the upper panels and 2$\rho_0$ in the lower panels) and two momenta (left and right). The $\bar K$ in matter feels attraction, that is seen by the shift of the peak of the spectral function to lower energies, while it develops a considerable width that is dominated by  the $S$-wave $\Lambda(1405)$ close to $\bar KN$ threshold.

With this figure we also indicate the importance of considering higher partial waves as well as taking into account the effect of finite temperature corrections for the determination of the properties of $\bar K$ in dense matter \cite{Tolos:2008di,Tolos:2006ny,Lutz:2007bh,Cabrera:2009qr,Cabrera:2014lca}. In particular, the knowledge of higher-partial waves beyond $S$-wave  as well as finite temperature corrections become fundamental for analyzing the results of  heavy-ion collisions (HICs) at beam energies below 2 AGeV \cite{Cassing:2003vz,Tolos:2003qj,Hartnack:2011cn}.

\section{Experiments and observations: heavy-ion collisions}
\label{hics}

A possible way to determine the antikaon optical potential (or self-energy) is to study the formation and propagation of antikaons in the dense medium created in nucleus-nucleus collisions for intermediate beam kinetic energies (GeV). The analysis of HICs for the determination of the properties of antikaons in dense matter requires, however, the use of transport models.  Transport models are he link between the experiments and the underlying physical processes, as they consider the production and propagation of all kind of particles, such as strange mesons (see Ref.~\cite{Hartnack:2011cn} for a review on strangeness production). Those models are based on the solution of Boltzmann-type semi-classical transport equations, that can be derived from the non-equilibrium quantum field theory.

The first transport calculations for antikaons were performed neglecting the finite width of the antikaon spectral function \cite{Cassing:1999es,Hartnack:2001zs}. Few years later, antikaon production was studied using off-shell dynamics with in-medium spectral functions in the Hadron-String-Dynamics (HSD) transport model \cite{Cassing:2003vz}. In this case,  the J\"ulich meson-exchange model \cite{Tolos:2000fj,Tolos:2002ud} was used for the  $\bar KN$ interaction. More recently, the in-medium effects in strangeness production in HICs at (sub-)-threshold energies of 1 - 2 A GeV based on the microscopic Parton-Hadron-String Dynamics (PHSD) transport approach have been studied, taking the in-medium antikaon properties  from the $\chi EFT$ scheme of Refs.~\cite{Tolos:2006ny,Cabrera:2014lca}. The manifestation of the medium effects  has been analyzed
for the strange mesons rapidity distributions, $p_T$-spectra as well as the polar and azimuthal angular distributions, directed and elliptic  flow in C$+$C, Ni$+$Ni, and Au$+$Au collisions. By comparing to experimental data from the KaoS, FOPI and HADES Collaborations, it was found that  the modifications of the strange meson properties in nuclear matter are necessary to explain the data in a consistent manner \cite{Song:2020clw}.

\section{Experiments and observations: neutron stars}
\label{neutronstars}

One of the possible scenarios inside the core of neutron stars is the appearance of  antikaons. The composition of matter in neutron stars is found by demanding equilibrium against weak interaction processes. Thus, considering neutron star matter made of neutrons, protons and electrons, the weak interaction transitions that take place are  
\begin{eqnarray}
n \rightarrow p e^- \bar \nu_e \nonumber \\
e^- p \rightarrow n \nu_e ,
\end{eqnarray} 
so that $\mu_p=\mu_n+\mu_e$ and $\rho_p=\rho_e$, with $\rho=\rho_p+\rho_n$. Nevertheless, if the chemical potential of the electron increases dramatically with density in the interior of neutron stars, it could become energetically more favourable to produce antikaons instead of electrons, via  
\begin{eqnarray}
n \leftrightarrow p + \bar{K} .
\end{eqnarray}
 For this to happen, the chemical potential of the electron for a given density should be larger than the effective mass of antikaons at that density, that is, $\mu_e > m^*_{\bar K}$. If this happens and due to the fact that antikaons are bosons, the phenomenon of kaon condensation would take place.

The possibility of kaon condensation has been debated since the pioneering work of Ref.~\cite{Kaplan:1986yq}.  The discussion is based on whether the mass of antikaons could be largely modified by the interaction with the surrounding nucleons.  Some phenomenological models tend to agree with this scenario. In particular, RMF models, that include or not other strange particles such as hyperons (see, for example, the recent works  \cite{Gupta:2013sna,Malik:2021nas,Muto:2021jms,Thapa:2021kfo}), seem to allow for this possibility. However, such large modifications in the mass of antikaons is not supported by $\chi$EFT approaches \cite{Ramos:1999ku,Tolos:2006ny,Cabrera:2014lca}.

\section{Summary}
\label{summary}

We have reviewed the status in the field of strange mesons, such as antikaons, in nuclei and neutron stars. We have analyzed the $\bar KN$ interaction, paying a special attention to  the role of the $\Lambda(1405)$. We have also  presented the theoretical efforts for determining the binding energy and width of the $\bar KNN$  bound state. We have then shown the properties of antikaons in dense nuclear matter, by displaying the $\bar K$ spectral function. We have finally discussed the propagation of strange mesons in HICs and the phenomenon of kaon condensation in neutron stars.

\section*{Acknowledgments}

L.T. acknowledges support from CEX2020-001058-M (Unidad de Excelencia ``Mar\'{\i}a de Maeztu"), PID2019-110165GB-I00 financed by the spanish MCIN/ AEI/10.13039/501100011033/, as well as by the EU STRONG-2020 project, under the program  H2020-INFRAIA-2018-1 grant agreement no. 824093, and the CRC-TR 211 'Strong-interaction matter under extreme conditions'- project Nr. 315477589 - TRR 211. 



%
 \bibliography{Bibliography}
%
%
%
%

\end{document}